%% file: main.tex
\documentclass{article}
\usepackage[T1]{fontenc}
\usepackage{spconf,amsmath,amssymb,graphicx,booktabs}
\usepackage[nocompress]{cite}
\usepackage[hidelinks]{hyperref}
\usepackage{etoolbox}
\usepackage{tikz}
\usetikzlibrary{arrows.meta}
\makeatletter
\patchcmd{\@maketitle}{\large \bf}{\fontsize{14}{16}\selectfont\bfseries}{}{}
\patchcmd{\@makefntext}{\hb@xt@1.8em}{\hb@xt@0pt}{}{}
\makeatother

\title{Shared Sentiment Evidence from Frozen Multimodal Models for Continuous Sentiment Prediction}
\name{Shuobo Wang$^{1}$, Ruiqi Chen$^{2}$, Shaobo Qiao$^{3}$, Shixu Yang$^{4}$, Feng Xu$^{1,*}$ \thanks{\fontsize{9}{11}\selectfont\raggedright *Corresponding author: Feng Xu\\Email: fengxu2@unsw.edu.au}}

\address{
$^{1}$UNSW Sydney, Australia \quad
$^{2}$University of Washington, USA\\
$^{3}$EULER AI, Australia \quad
$^{4}$SAIC Motor, China
}

\begin{document}
\ninept
\maketitle
\begin{abstract}
\input{sections/abstract}
\end{abstract}
\input{sections/introduction}


\input{sections/method}
\input{sections/experiments}
\input{sections/conclusion}
\clearpage
\bibliographystyle{IEEEbib-shortauthors}
\begingroup
\interlinepenalty=10000
\bibliography{references}
\endgroup
\end{document}

%% file: sections/abstract.tex
Continuous multimodal sentiment analysis estimates polarity and intensity from linguistic, acoustic, and visual cues that may disagree. Hard judgments compress sentiment alternatives, while learned fusion interfaces tie task adaptation to model representations. We introduce Shared Sentiment Evidence (SSE), an output-space interface with fixed view and sentiment-option semantics across frozen backbones. Four queries yield five-option distributions whose probabilities, entropy, and cross-view relations form a 28-dimensional vector for CatBoost regression. On CH-SIMS-v2 and CMU-MOSI, the Qwen-based interface improves on supervised joint-only readout. Matched Ridge and CatBoost controls assess the contributions of additional views, deterministic summaries, and nonlinear readout. With the same interface specification and fixed readout settings, separately fitted heads also improve on joint-only prediction for three additional frozen backbones. These results support learning continuous sentiment from reusable output coordinates.

%% file: sections/introduction.tex
\section{Introduction}
\label{sec:intro}
Continuous multimodal sentiment analysis maps linguistic, acoustic, and visual cues to a real-valued estimate of polarity and intensity. These cues can support different interpretations, with relative importance varying across clips. Continuous prediction requires both combining this evidence and learning how its strength relates to the annotation scale.

Task-specific multimodal sentiment analysis (MSA) learns representations and their interactions. TFN models cross-modal interactions, while MulT uses directional attention for unaligned streams~\cite{zadeh-etal-2017-tensor,tsai-etal-2019-multimodal}. MISA separates shared and modality-specific representations; Self-MM learns unimodal features through automatically generated auxiliary supervision~\cite{hazarika2020misa,yu2021selfmm}. Other methods organize fusion through language guidance, sentiment knowledge, and confidence-informed attention~\cite{zhang-etal-2023-learning-language,feng-etal-2024-knowledge,Jiang_2026_CVPR}. Recent works explore progressive fusion and semantic resonance/divergence with mixture-of-experts~\cite{li2026hpfn,sun2026dpmsa}. MEERA instead constructs evidence tokens and trains sparse adaptation branches around a frozen language model~\cite{Niu_2026}. These methods place task supervision in the representation, fusion, or adapter pathway; freezing a backbone does not itself remove the need to learn such an interface.

Multimodal large language models (MLLMs) offer further adaptation routes. MSE-Adapter and SentiLLM learn interfaces to language-model representations~\cite{yang2025mseadapter,chen2026sentillm}; prompt-guided teacher--student learning and in-context demonstrations exploit MLLM knowledge differently~\cite{zhang2025mmslf,wu2025sentimenticl}. The latter studies image--text sentiment classification, whereas our target is continuous audiovisual sentiment. MSACC explores contrastive zero-shot sentiment classification~\cite{liang2025msacc}. Unlike QLoRA-adapted hidden-state readout~\cite{wen2026beyond}, SSE learns from candidate output scores without hidden-state access or multimodal-backbone updates. At the output side, MPD offers hidden-state and calibrated class-score decoding strategies for multiple prompts~\cite{cheng2025mpd}. These routes distinguish learning an internal interface from learning on candidate scores. We study the latter across modality-conditioned views of the same ordered sentiment options.

Soft outputs preserve more than a winning option: distillation conveys relations among classes~\cite{hinton2015distilling}, while label distribution learning represents how multiple labels describe an instance~\cite{geng2016labeldistribution}. Our distributions are scores over prompted options, supervised by scalar sentiment annotations. Clips sharing a winning option can differ in adjacent-intensity support, and unimodal views can favor different polarities. However, hard decisions discard these alternatives, while fixed expectations compress them before learning their continuous-scale mapping. Joint-only learning retains distribution shape without explicitly exposing individual views, limiting the readout's access to modality-conditioned evidence.

To address this issue, we introduce \emph{Shared Sentiment Evidence (SSE)}. Text, audio, video, and joint queries populate fixed sentiment-option coordinates; a supervised readout learns their continuous-scale mapping. Coordinate definitions are shared across backbones, with a separate head for each backbone--dataset pair.

Our system uses Qwen2.5-Omni-7B~\cite{xu2025qwenomni} and CatBoost~\cite{prokhorenkova2018catboost}, without hidden-state access or backbone updates. Matched Ridge and CatBoost controls assess additional views and summaries. SSE improves on joint-only CatBoost on both datasets; fixed settings also support separate heads for Gemma, Phi, and Nemotron.

\begin{samepage}
\begin{list}{}{\setlength{\leftmargin}{\parindent}\setlength{\rightmargin}{0pt}\setlength{\topsep}{0pt}\setlength{\partopsep}{0pt}\setlength{\parsep}{0pt}}
\item[]
\noindent Our main contributions are:
\setlength{\leftmargini}{1em}
\setlength{\leftmarginii}{1em}
\begin{itemize}
\setlength{\itemsep}{0pt}\setlength{\parskip}{0pt}\setlength{\parsep}{0pt}
\item We formulate SSE, an output-space interface with fixed view, sentiment-option, and relation semantics across frozen multimodal backbones.
\item We instantiate SSE as a compact 28-D representation comprising four-view sentiment distributions and deterministic entropy and cross-view relation summaries, enabling continuous prediction from cached model outputs.
\item We establish SSE's value and cross-backbone reuse with matched controls isolating the effects of additional views, deterministic summaries, and nonlinear mapping; under fixed settings, SSE reduces MAE versus joint-only prediction across three additional backbones on both datasets.
\end{itemize}
\end{list}
\par
\end{samepage}

\begin{figure*}[t]
\centering
\includegraphics[width=\textwidth]{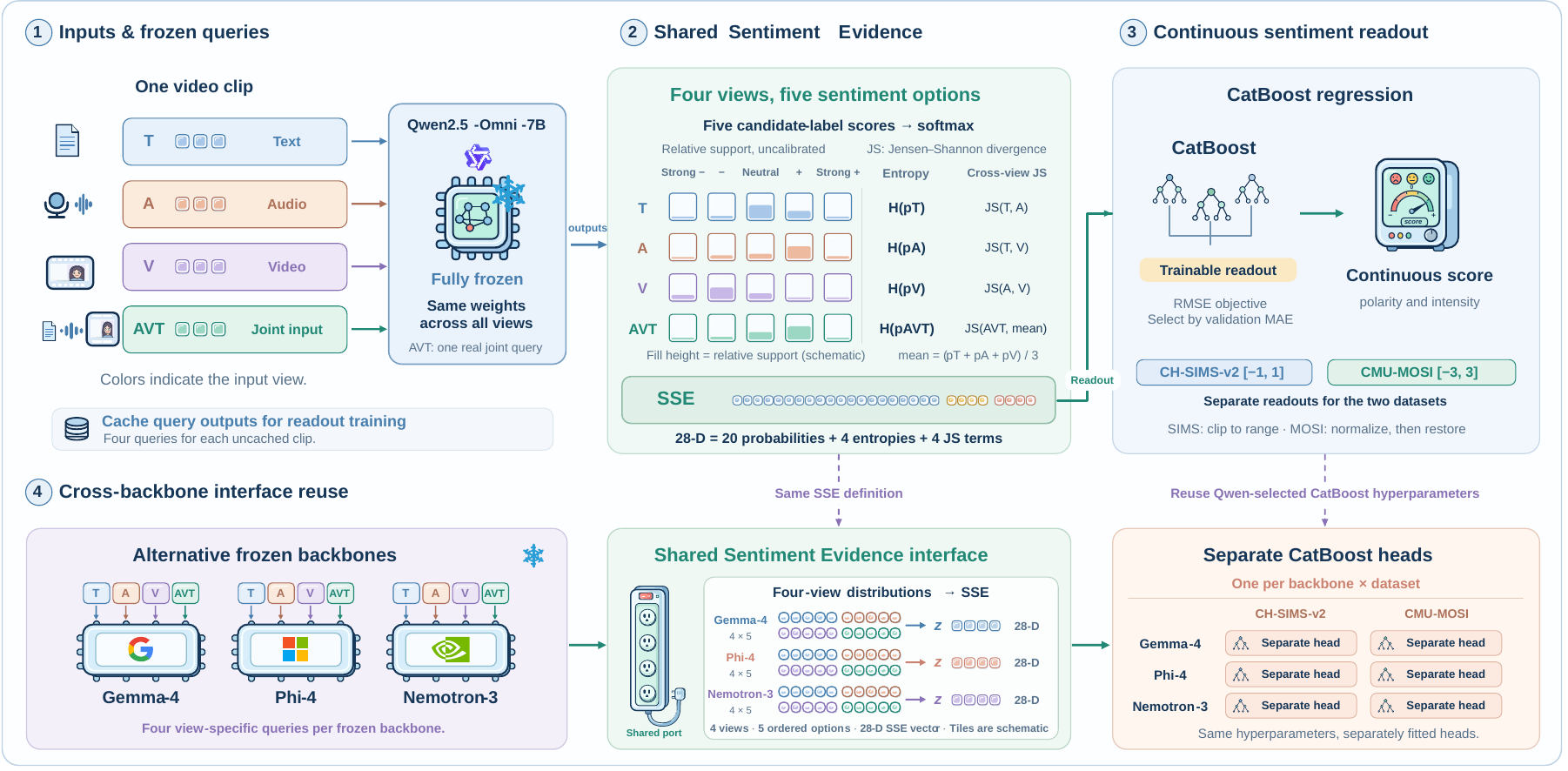}
\caption{Shared Sentiment Evidence for continuous readout across frozen multimodal backbones. Four view-conditioned sentiment distributions and their entropy and JS summaries form a 28-D output-space interface. Each backbone--dataset pair uses a separately fitted CatBoost head; probability-bar heights are schematic.}
\label{fig:method}
\end{figure*}

%% file: sections/method.tex
\section{Method}
\label{sec:method}
Figure~\ref{fig:method} presents SSE. A clip $x=(x_T,x_A,x_V)$ contains transcript, audio, and video, with sentiment label $y$. A supervised readout maps the frozen backbone's modality-conditioned distributions to the dataset's continuous scale.

\subsection{Frozen Multi-view Sentiment Queries}
\label{sec:queries}
Four views $\mathcal V=\{T,A,V,J\}$ query transcript, audio, video, and joint audio--video--text input using the same backbone weights. The joint query retains integrated judgment alongside unimodal judgments. We use Qwen's Thinker and each backbone's released media processors.

The prompt asks for the sentiment intensity expressed by the speaker, using digits 0--4 for strongly negative, negative, neutral, positive, and strongly positive. Let $\ell_{v,k}(x)$ denote the next-token logit or log-probability of candidate $k$ under view $v$. Normalization over these five candidates gives
\begin{equation}
p_{v,k}(x)=\frac{\exp\ell_{v,k}(x)}{\sum_{j=0}^{4}\exp\ell_{v,j}(x)},\quad k=0,\ldots,4.
\label{eq:prob}
\end{equation}
All five scores are extracted at the answer position using the backbone's tokenizer and chat template, without top-$k$ truncation. Each view satisfies $\sum_{k=0}^4 p_{v,k}=1$. These probabilities express relative option support, not empirically calibrated confidence~\cite{guo2017calibration}. Four queries supply 20 probabilities per uncached clip.

\subsection{Shared Sentiment Evidence}
\label{sec:evidence}
We retain the four distributions together with their entropy and Jensen--Shannon (JS) divergence~\cite{lin1991divergence}. For $m=(p+q)/2$,
\begin{align}
H(p)&=-\sum_k p_k\log p_k,\nonumber\\
\operatorname{JS}(p,q)&=\tfrac12\operatorname{KL}(p\Vert m)+\tfrac12\operatorname{KL}(q\Vert m).
\label{eq:js}
\end{align}
Natural logarithms are used throughout. With $\bar p=(p_T+p_A+p_V)/3$, the relation features are
\begin{equation}
\begin{split}
d=\big[&\operatorname{JS}(p_T,p_A),\operatorname{JS}(p_T,p_V),\\
&\operatorname{JS}(p_A,p_V),\operatorname{JS}(p_J,\bar p)\big].
\end{split}
\label{eq:relations}
\end{equation}
We construct the 28-dimensional SSE vector from 20 modality-conditioned sentiment probabilities, four entropy terms, and four cross-view JS terms:
\begin{equation}
z=\phi(x)=\big[p_T;p_A;p_V;p_J;\{H(p_v)\}_{v\in\mathcal V};d\big]\in\mathbb R^{28}.
\label{eq:features}
\end{equation}
The probabilities retain sentiment alternatives; entropy describes concentration, and JS exposes cross-view relations. These deterministic summaries provide nonlinear functions of the probabilities, not additional backbone evidence. The joint-to-mean term compares integrated judgment with unimodal consensus while retaining $p_J$ in $z$. Entropy lies in $[0,\log5]$ and JS in $[0,\log2]$, with fixed definitions across backbones.

\subsection{Lightweight Readout and Interface Reuse}
\label{sec:readout}
A CatBoost regressor $f_\theta$ learns $\hat y=f_\theta(z)$ from training annotations with the RMSE objective. CH-SIMS-v2 labels lie in $[-1,1]$.

\begin{samepage}
MOSI labels are divided by three for fitting, and predictions are multiplied by three for reporting. CH predictions are clipped to $[-1,1]$; MOSI regression uses continuous, unclipped predictions. The dataset-specific postprocessing is shared by the corresponding controls.
\par
\end{samepage}

For backbone $b$ and dataset $d$, $\hat y=f_{\theta_{b,d}}(\phi_b(x))$. SSE fixes coordinate definitions while $\phi_b(x)$ depends on the backbone's scores. We reuse the specification and selected CatBoost hyperparameters, fitting $\theta_{b,d}$ separately for each pair.

Only the readout learns task parameters; the backbone and modality adapters stay frozen. Cached distributions support readout fitting and feature ablations without repeated multimodal inference.

%% file: sections/experiments.tex
\section{Experiments}
\label{sec:experiments}
\subsection{Experimental Setup}
\label{sec:setup}
\begin{table}[t]
\input{tables/ch_main}
\par\vspace{4pt}
\input{tables/mosi_main}
\end{table}
\textbf{Data and metrics.} CH-SIMS-v2~\cite{liu2022chsims2} contains 2,722 training, 647 validation, and 1,034 test clips. The corresponding MOSI~\cite{zadeh2016sentimentintensity} split contains 1,284, 229, and 686 clips. MAE and Pearson correlation (Corr) use each dataset's original label scale. CH additionally reports Acc-2, weighted F1, Acc-3, and Acc-5. MOSI Acc-7 rounds clipped predictions and labels; Has0 binary metrics separate negative from nonnegative sentiment, while Non0 excludes zero labels. Accuracies and F1 are percentages. Evaluation follows the supplied benchmark splits; CH source videos occur across splits, whereas MOSI sources are disjoint.

\textbf{Main implementation.} Frozen Qwen2.5-Omni-7B runs in bfloat16 on an RTX 5090 with 16-kHz mono audio and 1-frame/s video capped at 200,704 pixels. For CMU-MOSI, grouped scaled dot-product attention (SDPA) batches independent visual sequences while retaining within-sequence attention, model weights, prompts, frame rate, and resolution. All four views use the same backend.

\textbf{Readout selection.} Qwen's readout family and hyperparameters are selected by validation MAE. We then fix the selected SSE--CatBoost configuration for test evaluation and additional backbones, fitting a separate head on each backbone--dataset training split.

The CatBoost search evaluates 12 combinations of depths $\{4,6,8\}$, learning rates $\{0.03,0.1\}$, and $L_2$ leaf regularization $\{3,10\}$, with 500 trees. Qwen SSE selects depth 6, rate 0.03, and regularization 3; joint-prob5 selects depth 4 under the same protocol. Ridge~\cite{hoerl1970ridge} standardizes training features and selects $\alpha$ by validation MAE from $\{0.01,\allowbreak0.1,\allowbreak1,\allowbreak10,\allowbreak100,\allowbreak1000\}$. Controls use the same caches, splits, targets, and evaluation, with selection conducted within each learner family.

\textbf{External comparisons.} An asterisk (*) marks our runs of released code under the stated adaptations; a dagger ($\dagger$) denotes published results. KuDA retains its released routing, and MSE-Adapter uses the shared official feature archive. For SentiLLM, we retain the released architecture and select checkpoints on the official validation split using Acc-5 on CH and MAE on MOSI; test IDs are reserved for final evaluation.

\textbf{Cross-backbone protocol.} Qwen2.5-Omni-7B~\cite{xu2025qwenomni} is the reference. Gemma 4 E4B-it~\cite{gemmateam2026gemma4}, Phi-4-multimodal-instruct~\cite{abouelenin2025phi4}, and Nemotron 3 Nano Omni~\cite{nvidia2026nemotron} reuse its option semantics, interface definition, and selected readout settings. Each backbone returns five candidate scores for the same four views. Phi receives ordered sampled images through frozen modality adapters; Gemma and Nemotron use their video interfaces. Nemotron uses NVFP4 weights and an FP8 KV cache. Joint5 and SSE use the same fixed depth-6 CatBoost configuration for these backbones, isolating the evidence interface without retuning.

\textbf{Joint-view comparators.} We use two joint-view controls. A fixed ordinal expectation maps the five joint-view probabilities to the full label range:
\begin{equation}
\hat y_{\mathrm{direct}}=s\sum_{k=0}^{4}c_kp_{J,k},\quad
c_k=\frac{k-2}{2},
\label{eq:direct}
\end{equation}
\begin{samepage}
where $s=1$ for CH-SIMS-v2 and $s=3$ for CMU-MOSI, giving the training-free baseline $c=(-1,-0.5,0,0.5,1)$. Joint-prob5 CatBoost uses only $p_J$ under the stated selection protocol. Comparing joint-prob5 with SSE tests the richer interface against supervised joint-only prediction; prob20 separates additional views from derived summaries. For statistical comparison, paired 95\% MAE intervals use 5,000 source-video bootstrap draws over 142 CH and 31 MOSI test groups. Whole-group resampling estimates clip-weighted MAE differences and quantifies test-sample uncertainty conditional on the fitted heads.
\par
\end{samepage}

\textbf{Evaluation design.} Tables~\ref{tab:ch} and~\ref{tab:mosi} evaluate continuous prediction; Table~\ref{tab:cross_backbone} holds option semantics and readout settings fixed across backbones; and Table~\ref{tab:ablation} uses matched Ridge and CatBoost controls. Together, these comparisons test predictive accuracy, interface reuse, downstream cost, and the separate effects of additional views, deterministic summaries, and nonlinear mapping.

\input{tables/cross_backbone}
\input{tables/ablation}
\subsection{Experimental Results and Analysis}
\label{sec:results}
\textbf{Continuous prediction.} On CH-SIMS-v2, the system reaches MAE 0.2483 and Corr 0.8157, with lower MAE than the reproduced external baselines in Table~\ref{tab:ch}; CICA reports 0.245. On MOSI, our 0.5660/0.8866 improves on the listed systems, including reproduced SentiLLM at 0.6650/0.8378 (Table~\ref{tab:mosi}). MOSI gains also appear in binary and seven-class metrics derived from the same predictions. CH rankings vary by metric, with KuDA leading Acc-5.

\textbf{Reuse across backbones.} SSE with CatBoost attains lower MAE than both Direct and supervised Joint5 for all four backbones on both datasets (Table~\ref{tab:cross_backbone}). Under identical depth-6 settings, Phi improves from 0.3301 to 0.3069 on CH and from 0.7780 to 0.6825 on MOSI; Gemma and Nemotron also improve on both datasets. These controls test the representation after supervised joint-score mapping is already available.
Across the six additional-backbone comparisons, paired SSE-minus-Joint5 intervals exclude zero for Phi on CH ($[-0.0343,-0.0123]$) and MOSI ($[-0.1349,-0.0506]$), and for Nemotron on CH ($[-0.0147,-0.0011]$). The other three include zero, while Corr point estimates improve in all six comparisons. Thus, all point estimates favor SSE, with the strongest paired evidence for Phi on both datasets and Nemotron on CH. Relative to Joint5, SSE reduces Qwen MAE by 11.1\% on CH and 4.2\% on MOSI; reductions across the six other cases span 1.5--12.3\%. Fixed option ordering, features, and readout settings test reuse of the evidence interface across backbone-specific scores.

\textbf{Downstream cost.} After caching Qwen outputs, its 12-setting CatBoost sweep takes 5.86 s on CH and 5.05 s on MOSI using 25 CPU threads. Each additional backbone's cache contains 26,408 view queries over both datasets; fixed-readout fitting and evaluation take approximately 2.7--3.0 s. Nemotron's full-cache stage takes 23.2 min, peaking at 31.2 GiB GPU memory. SSE uses four view queries per clip versus one for Direct and Joint5. This cost is incurred once during cache construction; learner selection and ablations reuse the probabilities, separating backbone inference from the seconds-scale CPU readout.

Nemotron has the lowest direct MAE on MOSI, whereas Qwen has the lowest MAE after readout on both datasets. This contrast shows that direct scores do not fully characterize the utility of a model's distributions for supervised prediction.

\subsection{Ablation Studies}
\label{sec:ablation}
Table~\ref{tab:ablation} varies the output representation and learner using the same frozen Qwen cache. Dataset-provided AVT features with Ridge form an independent conventional-feature control.

\textbf{Conventional features.} Dataset-provided AVT features with Ridge yield MAE 0.3097 on CH and 0.9993 on MOSI, compared with 0.3064 and 0.6203 for joint probabilities with Ridge. The compact joint output is competitive under this linear learner, especially on MOSI; the remaining controls vary the representation within the same Qwen cache.

\textbf{Joint-view readout.} Learning from the five joint probabilities improves fixed expectation conversion on both datasets. CatBoost further improves the same prob5 input over Ridge. Comparing prob5 and SSE using the same CatBoost family and validation-selection protocol tests the richer interface against supervised joint-only readout: MAE falls from 0.2794 to 0.2483 on CH and from 0.5910 to 0.5660 on MOSI, with paired source-video intervals excluding zero.

\textbf{Representation and relations.} Expanding the joint-view distribution to four modality-conditioned distributions consistently improves prediction under both readout families. With Ridge, MAE decreases from 0.3064 to 0.2828 on CH-SIMS-v2 and from 0.6203 to 0.5932 on MOSI. The same trend holds with CatBoost: four-view prob20 reduces MAE from 0.2794 to 0.2613 and from 0.5910 to 0.5681, respectively. Adding the deterministic entropy and cross-view JS coordinates further yields 0.2693/0.5921 with Ridge and 0.2483/0.5660 with CatBoost. MAE and Corr both improve at each step from joint prob5 to prob20 to SSE under either learner. The results indicate that multi-view probabilities provide the dominant shared gain, while the explicit summary coordinates offer an additional improvement that is more pronounced on CH-SIMS-v2.

\textbf{Nonlinear mapping.} Replacing Ridge with CatBoost on SSE lowers MAE from 0.2693 to 0.2483 on CH and from 0.5921 to 0.5660 on MOSI, while improving Corr on both. The controls support retaining modality-conditioned distributions and learning their task-scale mapping.

The summaries change how the learner accesses the existing probabilities. For Ridge, entropy and JS supply nonlinear basis functions of prob20; for finite-depth CatBoost, they permit direct splits on distribution concentration and cross-view disagreement. Their larger incremental gains on CH suggest dataset-dependent usefulness of this parameterization. Together, these controls distinguish the benefit of additional modality-conditioned evidence from the benefit of exposing its aggregate properties to the readout.

%% file: tables/ch_main.tex
\begingroup
\centering
\caption{CH-SIMS-v2 results. MAE uses $[-1,1]$; accuracy and F1 are percentages. Symbols follow Sec.~\ref{sec:setup}; dashes denote unreported metrics.}
\label{tab:ch}
\setlength{\tabcolsep}{1pt}
\renewcommand{\arraystretch}{1.0}
\begin{tabular*}{\columnwidth}{@{\extracolsep{\fill}}lcccccc@{}}
\toprule
Method & MAE\,$\downarrow$ & Corr\,$\uparrow$ & Acc-2\,$\uparrow$ & F1\,$\uparrow$ & Acc-3\,$\uparrow$ & Acc-5\,$\uparrow$\\
\midrule
KuDA$^*$~\cite{feng-etal-2024-knowledge} & 0.281 & 0.738 & 80.08 & 80.16 & 74.37 & \textbf{56.77}\\
MSE-Adapter$^*$~\cite{yang2025mseadapter} & 0.328 & 0.636 & 78.24 & 78.02 & 71.08 & 53.19\\
SentiLLM$^*$~\cite{chen2026sentillm} & 0.335 & 0.669 & 78.53 & 78.61 & 72.44 & 49.32\\
3WD-DRT$^\dagger$~\cite{Jiang_2026} & 0.276 & 0.764 & 83.05 & 84.01 & 75.86 & ---\\
CICA$^\dagger$~\cite{Jiang_2026_CVPR} & \textbf{0.245} & \textbf{0.842} & \textbf{85.98} & \textbf{85.89} & \textbf{80.56} & ---\\
MEERA$^\dagger$~\cite{Niu_2026} & 0.298 & 0.712 & 82.73 & 80.14 & 73.45 & 54.32\\
\midrule
\textbf{Ours} & 0.248 & 0.816 & 84.14 & 84.19 & 76.31 & 56.00\\
\bottomrule
\end{tabular*}\par
\endgroup

%% file: tables/mosi_main.tex
\begingroup
\centering
\caption{CMU-MOSI results. MAE uses $[-3,3]$; accuracy and F1 are percentages. Binary metrics list Has0 / Non0. Symbols follow Table~\ref{tab:ch}.}
\label{tab:mosi}
\setlength{\tabcolsep}{0.6pt}
\renewcommand{\arraystretch}{1.0}
\newcommand{\mosiAccTwo}[1]{\hspace*{-2pt}#1\hspace*{2pt}}
\begin{tabular*}{\columnwidth}{@{\extracolsep{\fill}}lccccc@{}}
\toprule
Method & MAE\,$\downarrow$ & Corr\,$\uparrow$ & Acc-7\,$\uparrow$ & \mosiAccTwo{Acc-2\,$\uparrow$} & F1\,$\uparrow$\\
\midrule
ALMT$^\dagger$~\cite{zhang-etal-2023-learning-language} & 0.683 & 0.805 & 49.42 & \mosiAccTwo{84.55\,/\,86.43} & 84.57\,/\,86.47\\
KuDA$^\dagger$~\cite{feng-etal-2024-knowledge} & 0.705 & 0.795 & 47.08 & \mosiAccTwo{84.40\,/\,86.43} & 84.48\,/\,86.46\\
3WD-DRT$^\dagger$~\cite{Jiang_2026} & 0.701 & 0.794 & 48.42 & \mosiAccTwo{83.41\,/\,86.51} & 85.28\,/\,87.08\\
CICA$^\dagger$~\cite{Jiang_2026_CVPR} & 0.630 & 0.855 & 49.56 & \mosiAccTwo{88.19\,/\,90.24} & 88.14\,/\,90.23\\
MEERA$^\dagger$~\cite{Niu_2026} & 0.672 & 0.812 & 46.83 & \mosiAccTwo{85.64\,/\,88.96} & 85.48\,/\,88.37\\
SentiLLM$^*$~\cite{chen2026sentillm} & 0.665 & 0.838 & 45.04 & \mosiAccTwo{86.30\,/\,88.57} & 86.25\,/\,88.57\\
\midrule
\textbf{Ours} & \textbf{0.566} & \textbf{0.887} & \textbf{50.73} & \mosiAccTwo{\textbf{88.34\,/\,90.55}} & \textbf{88.25\,/\,90.51}\\
\bottomrule
\end{tabular*}\par
\endgroup

%% file: tables/cross_backbone.tex
\begin{table}[!t]
\centering
\caption{Cross-backbone test MAE. Direct: fixed joint expectation; Joint5: joint prob5 + CatBoost; Ours: SSE + CatBoost. Readout heads are fitted separately.}
\label{tab:cross_backbone}
\setlength{\tabcolsep}{0.5pt}
\renewcommand{\arraystretch}{1.0}
\begin{tabular*}{\columnwidth}{@{\extracolsep{\fill}}lcccccc@{}}
\toprule
 & \multicolumn{3}{c}{CH-SIMS-v2} & \multicolumn{3}{c}{CMU-MOSI}\\
\cmidrule(lr){2-4}\cmidrule(lr){5-7}
Frozen backbone & Direct\,$\downarrow$ & Joint5\,$\downarrow$ & Ours\,$\downarrow$ & Direct\,$\downarrow$ & Joint5\,$\downarrow$ & Ours\,$\downarrow$\\
\midrule
Qwen2.5~\cite{xu2025qwenomni} & 0.374 & 0.279 & \textbf{0.248} & 0.815 & 0.591 & \textbf{0.566}\\
Gemma-4~\cite{gemmateam2026gemma4} & 0.333 & 0.289 & \textbf{0.282} & 0.837 & 0.652 & \textbf{0.619}\\
Phi-4~\cite{abouelenin2025phi4} & 0.390 & 0.330 & \textbf{0.307} & 0.887 & 0.778 & \textbf{0.683}\\
Nemotron-3~\cite{nvidia2026nemotron} & 0.363 & 0.260 & \textbf{0.252} & 0.784 & 0.601 & \textbf{0.591}\\
\bottomrule
\end{tabular*}
\end{table}

%% file: tables/ablation.tex
\begin{table}[!t]
\centering
\caption{Output-interface and readout ablations. All Qwen variants reuse frozen outputs; dataset-provided AVT features are an independent control.}
\label{tab:ablation}
\setlength{\tabcolsep}{1pt}
\renewcommand{\arraystretch}{1.0}
\begin{tabular*}{\columnwidth}{@{\extracolsep{\fill}}lcccc@{}}
\toprule
 & \multicolumn{2}{c}{CH-SIMS-v2} & \multicolumn{2}{c}{CMU-MOSI}\\
\cmidrule(lr){2-3}\cmidrule(lr){4-5}
Configuration & MAE\,$\downarrow$ & Corr\,$\uparrow$ & MAE\,$\downarrow$ & Corr\,$\uparrow$\\
\midrule
\begin{tabular}[c]{@{}l@{}}Direct joint-view\\expectation\end{tabular} & 0.374 & 0.715 & 0.815 & 0.865\\
\begin{tabular}[c]{@{}l@{}}Dataset-provided AVT\\features + Ridge\end{tabular} & 0.310 & 0.723 & 0.999 & 0.678\\
\midrule
Joint prob5 + Ridge & 0.306 & 0.732 & 0.620 & 0.870\\
Joint prob5 + CatBoost & 0.279 & 0.762 & 0.591 & 0.877\\
Four-view prob20 + Ridge & 0.283 & 0.779 & 0.593 & 0.878\\
Four-view prob20 + CatBoost & 0.261 & 0.802 & 0.568 & 0.885\\
SSE + Ridge & 0.269 & 0.795 & 0.592 & 0.880\\
\midrule
\textbf{SSE + CatBoost (Ours)} & \textbf{0.248} & \textbf{0.816} & \textbf{0.566} & \textbf{0.887}\\
\bottomrule
\end{tabular*}
\end{table}

%% file: sections/conclusion.tex
\section{Conclusion}
\label{sec:conclusion}
We presented SSE, a 28-D output-space interface that retains modality-conditioned sentiment distributions and their relations in fixed semantic coordinates. Matched linear and nonlinear controls support additional views and learned continuous-scale mapping. With three additional frozen backbones, the same specification and fixed settings improve on supervised joint-only prediction using separately fitted heads. SSE enables task adaptation through cached outputs while preserving coordinate definitions across models. Future work will examine missing modalities, cross-dataset calibration, and adaptive view acquisition, including query and annotation costs under changing modality availability.